\documentstyle [12pt]{article}
\begin{document} \title{The Stable Representation  of the algebra of functions
on the quantum group $SU_{q}(2)$.}
\author{S.V.Kozyrev \\ Steklov Mathematical Institute, Vavilov 42,}
\\
\maketitle \begin{abstract} An operation of a coproduct of representations
of a bialgebra is defined.  The coproduct operation for representations
of the Hopf algebra  of functions on the quantum group $SU_{q}(2)$
is investigated.  A notion of a
stable representation $\Pi$ is introdused.
This means that the representation $\Pi$ is invariant under coproduct
by arbitrary representation.  Algebra  of functions on the quantum group
$SU_{q}(2)$ in the representation $\Pi$ is a factor of a type
II$_{\infty}$.  Formula for the trace in the representation $\Pi$ is given .
The invariant integral of Woronovich on $SU_{q}(2)$ will take the form
$\int f d\mu  = tr(f cc^{*})$.
\end{abstract}
\newpage
\section{ Introduction . }

For representations of algebras  is well known such operation as a direct
sum. If an algebra possesses some additional structures  one can consider
additional operations for representations of this algebra.

In this paper
we define an operation that we call coproduct for representations
of a bialgebra and investigate its properties.
 The bialgebra $A$ is an associative algebra with additional
operation of a coproduct. The coproduct is a linear map $\Delta:A\to A\otimes
A$
wich is compatible with multiplication in the algebra A. This means that
$\Delta ab=\Delta a\Delta b$ where multiplication in $A\otimes A$ is defined
according to the formula $(a\otimes b)(c\otimes d)=ac\otimes bd;a,b,c,d\in A$.
The operation of the coproduct must be coassociative. This means that the
following condition does hold
$$(\Delta\otimes Id)\Delta=(Id\otimes\Delta)\Delta.$$
Let $S$ and $T$ be representations of an associative algebra $A$ which
have a bialgebra structure with the coproduct $\Delta$.
The coproduct $ST$ of representations $S$ and $T$ we will define according
to the formula  $ST(a)=S\otimes T(\Delta a); a\in A$.
Using associativity and distributivity of tensor product and coassociativity
of coproduct it is easy to prove the following proposition.

{\bf Proposition 1.}{\it  The operation of coproduct of representations
of bialgebra $A$ have following properties of associativity and distributivity:
$$S(TP)=(ST)P$$
$$S(T\oplus P)=ST\oplus SP$$
$$(S\oplus T)P=ST\oplus SP$$
Here $S,T,P$ are representations of bialgebra $A$.
}

In this paper we will
investigate coproducts of representations  for the algebra of functions
on the quantum group $SU_{q}(2)$.

The quantum group $SU_{q}(2)$ is an object of a great interest since 1986.
This is a Hopf algebra or bialgebra with some extra structures.
The Hopf algebra $A$ of functions on the $SU_{q}(2)$  is defined for real $q$,
we take $0<|q|<1$,
 and as algebra is generated by elements $a,b,c,d$ with
 involutive antihomomorphism $*$:

$\begin{array}{c}
a^{*}=d \\ b^{*}=-q c
\end{array}$

Commutational relations for
  algebra $A$ has the form:

$\begin{array}{cccc}

ab=qba & ac=qca & ad-qcb=1 & bc=cb \\
cd=qdc & bd=qdb & da-q^{-1}bc=1

\end{array}$

The Hopf algebra  $A$ has the coproduct structure.
This coproduct has the following form:
$$\Delta :A\to A\otimes A$$
$$\Delta g^{i}_{j}=\sum_{k=0,1}g^{i}_{k}\otimes g^{k}_{j}$$
where $(g^{i}_{j})=\left(\begin{array}{cc}a&b\\c&d\end{array}\right)$.

More complete information on quantum groups one can find for example
in papers [1],[2],[3]. The classification of representations of $SU_{q}(2)$
enveloping algebra one can find in the work [4].

Classification of irredusible *-representations of
$A$ was given by Vaxman and Soibelman in the paper [5] and has
the following form:
exist two continuous families of irredusible *-representations.

1) Family $\rho_{\phi}$:
\begin{equation}
\left(\begin{array}{cc}a&b\\c&d\end{array}\right)=
\left(\begin{array}{cc}\exp(i \phi)&0\\0&\exp(-i \phi)\end{array}\right)
\end{equation}
realized in one-dimensional Hilbert space.

2) Family $\pi_{\theta}$: Representations of this family are realised in
Hilbert space $H_{\pi_{\theta}}$ with orthonormal basis
$\{ e_{m}\} $ where $ m $  runs from $ 0 $ to $ \infty$
according to the following formulas:
\begin{equation}
a e_{m}=\sqrt{1-q^{2m}}e_{m-1} ;  a e_{0}=0
\end{equation}
\begin{equation}
d e_{m}=\sqrt{1-q^{2(m+1)}}e_{m+1}
\end{equation}
\begin{equation}
c e_{m}=\exp(i \theta) q^{m} e_{m}
\end{equation}
\begin{equation}
b e_{m}=-\exp(-i \theta) q^{m+1} e_{m}
\end{equation}
An operation of coproduct for *-representations of
$A$  will take the following form.
If $\xi$ and $\eta$ are *-representations of $A$
in correspondent Hilbert spaces $H_{\xi}$ and $H_{\eta}$ we define
a *-representation $\xi \eta$ of $A$ in the Hilbert space
$H_{\xi}\otimes H_{\eta}$
according to the following formula.

\begin{equation}
\left(\begin{array}{cc}a&b\\c&d\end{array}\right)_{\xi \eta}=
\left(\begin{array}{cc}a&b\\c&d\end{array}\right)_{\xi}\otimes
\left(\begin{array}{cc}a&b\\c&d\end{array}\right)_{\eta}
\end{equation}

We define a *-representation $\Pi=\pi_{\phi}\pi_{\psi}\pi_{\theta}$.
We will show that
the class of unitary equivalence of $\Pi$ does not depend on $\phi$,$\psi$ and
$\theta$. This representation has a remarkable property: class of unitary
equivalence of this representation is invariant under coproduct by
arbitrary representation.
We have $\Pi \xi=\xi \Pi=\Pi$
where $"="$ means unitary equivalence and $\xi$ is arbitrary representation.
  According to this property we will
call the representation $\Pi$ stable.

A weak closure of $A$ in the representation $\Pi$
is a factor of a type II${}_{\infty}$. A trace on this factor is connected with
the invariant integral of Woronowich on the quantum group
 $SU_{q}(2)$  by the formula $\int f d\mu  = tr(f cc^{*})$.

\section{ Some properties of representations of  $A$.}
Let us consider some properties of *-representations of $A$.
Let $\xi$ be a *-representations of $A$  in Hilbert space $H_{\xi}$.
For the polynom $f\in A$ we denote $f_{\xi}$ this polynom
in the representation $\xi$. We will not indicate the index $\xi$ if
the representation is clear from the context.

For the following lemma author is grateful to I.V.Volovich.

{\bf  Lemma 1.  }{\it  Every *-representation $\xi$ of algebra $A$
is realised by bounded operators. }

{\bf Proof}

Let us consider the commutational relation $$da-q^{-1}bc=1.$$
This condition is equivalent to $$a^{*}a+c^{*}c=1.$$
Therefore for arbitrary vector $x$ in the Hilbert space $H_{\xi}$
we will have the following.
$$(x,x)=(x,(a^{*}a+c^{*}c)x)=(ax,ax)+(cx,cx)$$
Thus operators $a$ and $c$ are bounded. It is easy to see that operators
$d=a^{*}$ and $b=-qc^{*}$ are also bounded. This proves the lemma.

The crucial fact in our considerations is the following.

{\bf  Theorem 1.  }{\it  Every *-representation $\xi$ of algebra
$A$ possesses an expansion in the direct sum
$\xi=\xi_{F}\oplus\xi_{G}$ of *-representations
in orthogonal subspases $F_{\xi}$ and $G_{\xi}$ with the following properties.
In $F_{\xi}$ we have $Kerc={0}$ and
$F_{\xi}=\oplus_{m=0}^{\infty} d_{\xi}^{m} Ker a_{\xi}$.
In $G_{\xi}$ we have $c=0$ and $a$ is unitary operator.}

{\bf  Proof}

Let us denote $F_{0}=Ker a_{\xi}$ and
$F_{m}=d_{\xi}^{m} Ker a_{\xi}; m=0,1,2,...$.
Let us prove that
$F_{m}$ are mutually orthogonal closed subspaces of $H$.
Let $x=d^{m}x_{0}\in F_{m}$ and $y=d^{n}y_{0}\in F_{n}$ for $m>n$.
We have $(x,y)=(d^{m}x_{0},d^{n}y_{0})=(x_{0},a^{m}d^{n}y_{0})=0$ due to
commutational relations and the condition $ay_{0}=0$.

Let us prove that $F_{m}$ are closed subspaces. It is easy to see that
$F_{0}=Ker a$ is closed. Let $F_{m}$ be closed. We will prove that
$F_{m+1}$ is closed. Let us consider a vector $x=dx_{0}\in F_{m+1}$
for $x_{0}\in F_{m}$.

Let us prove that $ax=adx_{0}=(1-q^{2(m+1)})x_{0}$.
We have for every $x_{0}\in F_{m},x_{0}=d^{m}z$ for $z\in F_{0}$
the condition $adx_{0}=(1+qbc)x_{0}=(1+qbc)d^{m}z=d^{m}(1+q^{2m+1}bc)z$.
Because $z\in F_{0}$ we have $az=0$ and $daz=(1+q^{-1}bc)z=0$.
Therefore $bcz=-qz$. Thus $adx_{0}=(1-q^{2(m+1)})d^{m}z=(1-q^{2(m+1)})x_{0}$.

Thus for every $x_{0}\in F_{m}$
we have a vector $x_{1}=\frac{1}{1-q^{2(m+1)}}dx_{0}\in F_{m+1}$ such that
$ax_{1}=x_{0}$. Because $Ker a=F_{0}$ is orthogonal to $F_{m+1}$
we have that preimage of $F_{m}$  with respect to the bounded operator
$a$ is an orthogonal sum $F_{m+1}\oplus F_{0}$.
Because preimage is closed and a closed subspace
$Ker a=F_{0}$ is orthogonal to $F_{m+1}$ we have proved that
$F_{m+1}$ is closed.

Let us define $F_{\xi}=\oplus_{m=0}^{\infty}F_{m}$.
$F_{\xi}$ is a closed subspace of $H$
and its orthogonal complement
$G_{\xi}=\{x\in H:(x,y)=0\forall y\in F\}$ is a closed
subspace of $H$.

We have in $F_{\xi}$ and $G_{\xi}$ a subrepresentations
of $\xi$
due to the facts
that $d$ maps $F_{m}$ into $F_{m+1}$, $a$ maps $F_{m}$ into $F_{m-1}$,
$c$ and $b$ maps $F_{m}$ into $F_{m}$ thus $ \forall f \in A$
 it maps $F_{\xi}$ into $F_{\xi}$.

To prove that
in $F_{\xi}$ we have $Kerc={0}$ it sufficient to prove that in $F_{0}$
we have $Kerc=0$. This proves as follows.
Let $x\in Kerc\bigcap F_{\xi}$.
We  have $x=\sum_{m=0}^{\infty}x_{m}, x_{m}\in F_{m}$. Due to commutational
relations we have $cx_{m}\in F_{m}$. Therefore exist $m$ with
$x_{m}\ne 0$ and $cx_{m}=0$. We have $x_{m}=d^{m}y$ for $y\in F_{0}$.
Taking nonzero vector $z=a^{m}x_{m}\in F_{0}$ we get $cz=q^{-m}a^{m}cx_{m}=0$.
We have proved that if $Kerc\bigcap F_{\xi}$ is nonzero subspace then
$Kerc\bigcap F_{0}$  is nonzero subspace.

For $x\in Kerc\bigcap F_{0}$  we   have $dax=x$ and $dax=0$. Thus $x=0$ and
$Kerc\bigcap F_{\xi}$ is zero subspace.

Let us prove that $c=0$ in $G_{\xi}$. Thus $a$ in $G_{\xi}$ will be unitary.
Let $c\ne 0$ in $G_{\xi}$.
Consider the following operator in $G$:
$$T=c^{*}c$$
If we denote the spectrum of the operator $T$ as $\sigma(T)$ we have
\begin{equation}\label{spect}
\sup_{\lambda \in \sigma(T)}\lambda=\sup_{x\in G, ||x||=1}(Tx,x)=
\sup_{x\in G, ||x||=1}(cx,cx)=||c||_{G}^{2}>0
\end{equation}

because  $c_{\xi}\ne 0$. $T$ is a bounded selfajoint operator
and its spectrum consists of discrete and continuous parts.
Consider the following cases:

1) Let $T$ have nonzero eigenvalues. All these eigenvalues are positive.
Consider an eigenvalue $\lambda$:
$\lambda> q^{2}\sup_{\mu \in \sigma_{d}(T)}\mu$;   where $\sigma_{d}(T)$
is discrete spectrum.  Let $x$ be an eigenvector: $Tx=\lambda x$.
Because of
$Ta=q^{-2}aT$
we have
$Ta x=q^{-2}\lambda a x$.
In $G$ we have $Ker a =\{0\}$. This means that $a x \ne 0$ and it is an
eigenvector with eigenvalue $q^{-2}\lambda$. We get a contradiction.

2)  Let nonzero spectrum of $T$ be purely continuous. Let $\lambda$ be
a maximal element of spectrum and let $\mu$ be an element of spectrum
such that $q^{2}\lambda<\mu \le \lambda$. We have $q^{-2}\mu$ does not
lie in spectrum.

For an element $\mu$ of continuous spectrum there exists sequence
$x_{k}, ||x_{k}||\ge 1 : ||(T-\mu)x_{k}||$ goes to zero. Then
$||(T-q^{-2}\mu)a x_{k}||=||q^{-2}a(T-\mu)x_{k}||$ goes to zero.
We have $q^{-2}\mu\in \sigma(T)$ if there is no
such subsequence $x_{k_{n}}: ||a x_{k_{n}}||$ goes to zero.

Taking the subsequense   $x_{k_{n}}$ we get the following.
Because $d$
is a bounded  operator  we have $||d a x_{k_{n}}||$ goes to zero.
Using commutational relations we have $ d a = 1-c^{*} c=1-T$. We get
$$(T-\mu)x_{k_{n}} \to 0: $$
$$(T-1)x_{k_{n}} \to 0 ;$$
for $||x_{k_{n}}|| \ge 1$ that means that $\mu=1$. This means that $1$ is
an isolated point of spectrum of a bounded selfajoint operator. This means
that $1$ is eigenvalue of $T$ and we get a contradiction.

We have shown that the operator $T$ has not a nonzero spectrum. Using equation
(\ref{spect}) we get the condition that $cx=0  \forall x\in G_{\xi}$.
The Theorem 1 is proved.

{\bf Corollary 1. }{\it  The subspace $G_{\xi}$ of the Theorem 1 coinsides
with $Kerc_{\xi}$.}

This theorem gives us a simple parametrization of all *-representations
of
algebra of functions
on  $SU_{q}(2)$.
*-Representations are parametrized by a couple of unitary operators
consisting of $c$ on $Kera$ and $a$ on $Kerc$.

Let us built a canonical orthonormal basis in the closed subspace $F$.
Let $\{e_{M}\}$ be an orthonormal basis in $F_{0}=Kera$ where an operator
$c$ acts as follows
$$ce_{M}=\sum_{N}c_{M}^{N}e_{N}$$ where $c_{M}^{N}$ are matrix elements
of unitary operator. In $F$ we have an orthonormal basis $\{e_{mM},m$ runs
from $0$ to $\infty\}$ where
$$e_{mM}=\frac{1}{(1-q^{2})^{\frac{m}{2}}\sqrt{[m]_{q^{2}}!}}d^{m}e_{M}.$$
It is easy to prove the following Lemma.

{\bf Lemma 2.  }{\it    In the orthonormal basis $\{e_{mM}\}$ operators of
algebra of functions on $SU_{q}(2)$ acts in the following way:

\begin{equation}
a e_{mM}=\sqrt{1-q^{2m}}e_{m-1,M} ; a e_{0M}=0
\end{equation}
\begin{equation}
d e_{mM}=\sqrt{1-q^{2(m+1)}}e_{m+1,M}
\end{equation}
\begin{equation}
c e_{mM}=q^{m}\sum_{N}c_{M}^{N} e_{mN}
\end{equation}
\begin{equation}
b e_{mM}=-q^{m+1}\sum_{N}c_{M}^{*N} e_{mN}
\end{equation}
 Here $c_{M}^{N}$ and $c_{M}^{*N}$ are
matrix elements of mutually conjugated unitary operators.}

For definitions of the next proposition see book [8].

{\bf Proposition 2.  }{\it    The subrepresentation of the representation
$\xi$ of $A$ in the subspace $F_{\xi}$  of the Proposition 1 is factorial.
This means that $A$ in this subrepresentation is a factor of a type
I$_{\infty}$
for irredusible subrepresentations and of a type  II$_{\infty}$  for
redusible subrepresentations. Using the basis of the Lemma 2 one can built
a semifinite trace on this factor according to the formula

$$tr(f)=\sum_{k=0}^{\infty} f_{k0}^{k0}$$
where $f^{mM}_{nM}=(e_{mM},fe_{nN})$ for $f\in A$. The notion $f_{k0}^{k0}$
means that  indexes $M$ and $N$
takes a value $0$.       }

{\bf Proof}

Let us prove that the center of $A$  contains only identity operator.
Let us first prove that the operator
$W$ commuting with $a$ and $d$ will have matrix elements
\begin{equation}\label{w}
W_{mM}^{nN}=(e_{nN},W e_{mM})=\delta_{m}^{n} W_{M}^{N}.
\end{equation}
 From the condition $Wa=aW$ it follows that
\begin{equation}\label{ww}
W_{mM}^{n+1,N}\sqrt{1-q^{2(n+1)}}= W_{m-1,M}^{nN}\sqrt{1-q^{2m}}.
\end{equation}
 From the formula (\ref{ww}) it follows that $W_{mM}^{n+1,N}=0$
for arbitrary $n$.
Using the formula (\ref{ww}) we get the condition
\begin{equation}\label{www}
W_{m+1,M}^{n+1,N}= W_{mM}^{nN}\sqrt{\frac{1-q^{2(m+1)}}{1-q^{2(n+1)}}}.
\end{equation}
Using the formula (\ref{www}) we get the condition $W_{mM}^{nN}=0$
for $n>m$.

Analogously from the condition $Wd=dW$
one can get the condition $W_{mM}^{nN}=0$
for $n<m$.

We have got that $W_{mM}^{nN}$  can be nonzero only for $m=n$.
The formula (\ref{www}) in this case takes the form
$W_{m+1,M}^{n+1,N}=W_{mM}^{nN}.$
We have proved the condition
$$W_{mM}^{nN}=\delta_{m}^{n} W_{M}^{N}. $$

Let us prove that a weak closure of $A$ in the representation considered
is a factor. Let $W$ be an element of the center of the
 weak closure of $A$ in the representation considered.

$W$ is a weakly convergent series on monoms $a^{k}b^{i}c^{j}$ and
$d^{k}b^{i}c^{j}$.
The matrix element of the monom  $a^{k}b^{i}c^{j}$  is proportional to
$\delta_{m}^{n+k}$. Analogously the matrix element of the monom
$d^{k}b^{i}c^{j}$ is proportional to $\delta_{m+k}^{n}$.  Therefore the
series $W$ can include
 only monoms of a kind $b^{i}c^{j}$.
 But the matrix element of the monom $b^{i}c^{j}$  is proportional to
 $\delta_{m}^{n}q^{n(i+j)}$.
 Using (\ref{w}) we have $i=j=0$ and $W$ is proportional to $1$.
 We have proved  that a weak closure of $A$
in the representation considered is a factor.

For irredusible representation
this factor will have a type I$_{\infty}$.
Let us prove that for arbitrary representation
considering factor does not admit a faithful trace $Tr$ of a
finite type. For this trace we would have $Tr(ad-da)=(q-q^{-1})Tr(bc)=0$.
This means that $Tr(cc^{*})=0$ and we get a contradiction.
Let us propose a formula for semifinite trace $tr$ on a considering factor.
This will mean that this factor will have a type  II${}_{\infty}$
for arbitrary redusible representation in the Hilbert space $F_{\xi}$.
\begin{equation}\label{tr}
tr(f)=\sum_{k=0}^{\infty} f_{k0}^{k0}
\end{equation}

Let  us prove that the formula (\ref{tr}) defines a trace.
It is sufficient to prove that for arbitrary $f,g\in A$
if $tr(fg)$ is defined then $tr(gf)$ will be defined and
the following condition will holds:

\begin{equation}\label{trcondition}
tr(fg)=tr(gf)
\end{equation}
It sufficient to prove the formula (\ref{trcondition}) for arbitrary
monoms $f$ and $g$. Taking monom $f=a^{k}b^{i}c^{j}$ we should prove
the formula (\ref{trcondition}) only for  $g=d^{k}b^{m}c^{n},\forall m,n$
because for another monoms $g$ we have $tr(fg)=0$ due to the formula (\ref{tr})
and Lemma 2.

We have to prove the formula (\ref{trcondition}) for $f=a^{k}b^{i}c^{j}$ and
$g=d^{k}b^{m}c^{n}$. Values  $tr(fg)$ and $tr(gf)$ both are defined for
$i+j+m+n>0$. In this case we have
$$tr(fg)=tr(a^{k}b^{i}c^{j}d^{k}b^{m}c^{n})=
q^{k(i+j)}tr(b^{i}c^{j}a^{k}d^{k}b^{m}c^{n})$$
and
$$tr(gf)=tr(d^{k}b^{m}c^{n}a^{k}b^{i}c^{j})=
q^{k(i+j)}tr(d^{k}b^{m}c^{n}b^{i}c^{j}a^{k})=
q^{k(i+j)}tr(a^{k}d^{k}b^{m}c^{n}b^{i}c^{j}).$$
We have used commutational relations and the fact that according to Lemma 2
the operator $a^{k}$ acts only on the first index. Therefore according to the
definition of trace (\ref{tr}) we can made transposition in the last formula.

To finish the proof we only have to note that due to commutational relations
$a^{k}d^{k}$ is a polynom on $b$ and $c$ and $b$ and $c$  are commute.
We have proved the Proposition 2.
\section{Coproducts of representations of $A$.}

Let us investigate the behavior of representations of $A$
under coproduct. Let $\xi$,$\eta$ be representations of $A$ and $\xi \eta$
be their coproduct.

\begin{equation}
\left(\begin{array}{cc}a&b\\c&d\end{array}\right)_{\xi \eta}=
\left(\begin{array}{cc}a&b\\c&d\end{array}\right)_{\xi}\otimes
\left(\begin{array}{cc}a&b\\c&d\end{array}\right)_{\eta}
\end{equation}

Let representation $\xi=\xi_{F}\oplus\xi_{G}$
be realized in Hilbert space $H_{\xi}$.
According to  Theorem 1 and Lemma 2 this Hilbert space has
an orthonormal basis
$\{e_{mM},m$ runs from $0$ to $\infty;e_{\alpha}\}$ where $\{e_{0M}\}$ is
an orthonormal basis
in $Ker a_{\xi}$
and $\{e_{\alpha}\}$ is an orthonormal basis in $G=Kerc_{\xi}$
and  operators of $A$ acts in
the following way:

\begin{equation}
a e_{mM}=\sqrt{1-q^{2m}}e_{m-1,M} ; a e_{0M}=0
\end{equation}
\begin{equation}
d e_{mM}=\sqrt{1-q^{2(m+1)}}e_{m+1,M}
\end{equation}
\begin{equation}
c e_{mM}= q^{m} \sum_{N} c_{M}^{N} e_{mN}
\end{equation}
\begin{equation}
a e_{\alpha}=\sum_{\beta}a_{\alpha}^{\beta} e_{\beta}
\end{equation}
\begin{equation}
c e_{\alpha}=0
\end{equation}
Where $c_{M}^{N}$ and $a_{\alpha}^{\beta}$ are matrixes of unitary operators.
Here we use notions of Theorem 1 and Lemma 2.

Analogously  representation $\eta=\eta_{F}\oplus\eta_{G}$
is realized in Hilbert space $H_{\eta}$
with an orthonormal basis
$\{e_{nN},n$ runs from $0$ to $\infty;e_{\gamma}\}$ where $\{e_{0N}\}$
is an orthonormal basis
in $Ker a_{\eta}$
and $\{e_{\gamma}\}$ is an orthonormal basis in $G=Kerc_{\eta}$
and an operators of $A$ acts in
the following way:

\begin{equation}
a e_{nN}=\sqrt{1-q^{2n}}e_{n-1,N} ; a e_{0N}=0
\end{equation}
\begin{equation}
d e_{nN}=\sqrt{1-q^{2(n+1)}}e_{n+1,N}
\end{equation}
\begin{equation}
c e_{nN}= q^{n} \sum_{K} c_{N}^{K} e_{nK}
\end{equation}
\begin{equation}
a e_{\gamma}=\sum_{\delta} a_{\gamma}^{\delta} e_{\delta}
\end{equation}
\begin{equation}
c e_{\gamma}=0
\end{equation}
Where $c_{N}^{K}$ and $a_{\gamma}^{\delta}$ are matrixes of unitary operators.

We have an important lemma.

{\bf  Lemma  3.}  {\it  For *-representations $\xi$ and $\eta$ of
$A$
we have $Ker c_{\xi \eta}=Ker c_{\xi}\otimes Ker c_{\eta}$}

{\bf Proof}

Representation $\xi \eta$ will be realized in the Hilbert space
$H_{\xi}\otimes H_{\eta}$.

An arbitrary
$x\in H_{\xi}\otimes H_{\eta}$ have the following form.

$$x=\sum_{m,n=0;M,N}^{\infty} x^{mnMN} e_{mM}\otimes e_{nN}
+\sum_{m=0;M,\beta}^{\infty} x^{mM\beta} e_{mM}\otimes e_{\beta}+$$
$$+\sum_{n=0;N,\alpha}^{\infty} x^{\alpha nN} e_{\alpha}\otimes e_{nN}+
\sum_{\alpha\beta} x^{\alpha\beta} e_{\alpha}\otimes e_{\beta}$$
Let the following condition holds:$x\in Ker c_{\xi \eta}$.

By acting
$c_{\xi \eta}=c_{\xi}\otimes a_{\eta}+d_{\xi}\otimes c_{\eta}$
on the first term of $x$ one gets the following.
$$
\sum_{m,n=0,K}^{\infty} x^{m,n+1,MN} q^{m} c_{M}^{K} \sqrt{1-q^{2(n+1)}}
e_{mK}\otimes e_{nN}
+$$
\begin{equation}+ \sum_{m=1,n=0,K}^{\infty} x^{m-1,n,MN} q^{n}
\sqrt{1-q^{2m}} c_{N}^{K}e_{mM}\otimes e_{nK} =0
\label{Kerc}\end{equation}
Therefore we have $x^{0,n+1,MN}=0 \forall n$. Iterating we have $x^{mnMN}=0 $
  for $n>m$. From the formula   (\ref{Kerc}) it  follows that
\begin{equation}
x^{m+1,n+1,MN}=-q^{n-m-1}\sqrt{\frac{1-q^{2(m+1)}}{1-q^{2(n+1)}}}
\sum_{K,L} c_{K}^{*M}c_{L}^{N}x^{mnKL}
\end{equation}
for $m+1\ge n$.
Because $|q^{n-m-1}|\ge 1$ we have $|x^{m+k,n+k,MN}|\to\infty$
for $k\to\infty$. This means that for $x\in Ker c_{\xi\eta}$ we have
$x^{mnMN}=0$. Analogously we have  for $x\in Ker c_{\xi\eta}$
 conditions $x^{mM\beta}=x^{\alpha nN}=0$ and $x^{\alpha\beta}$ are arbitrary.
This proves the Lemma 3.

Our aim is to construct the coproduct $\xi\eta$. According to Proposition 1
$$\xi\eta=\xi_{F}\eta_{F}\oplus\xi_{F}\eta_{G}\oplus\xi_{G}\eta_{F}
\oplus\xi_{G}\eta_{G}.$$
Therefore it sufficient to prove following propositions.

{\bf Proposition 3. }{\it
Representation $\xi_{F}\eta_{F}$ is realised in the Hilbert space
$F_{\xi}\otimes F_{\eta}$. This Hilbert space possess the orthonormal basis
$\{e_{mnMN},m=0,1,2...;n$~is~integer~$\}$. In this basis
representation  $\xi_{F}\eta_{F}$
is defined according to the following formulas:
\begin{equation}
a e_{mnMN}=\sqrt{1-q^{2m}}e_{m-1,nMN} ; a e_{0nMN}=0
\end{equation}
\begin{equation}
d e_{mnMN}=\sqrt{1-q^{2(m+1)}}e_{m+1,nMN}
\end{equation}
\begin{equation}
c e_{mnMN}= q^{m} e_{m,n+1,MN}
\end{equation}
\begin{equation}
b e_{mnMN}=- q^{m+1} e_{m,n-1,MN}
\end{equation}
}

{\bf Proof}

According to Lemma 3 we have $Kerc_{\xi_{F}\eta_{F}}=0$.
Therefore this representation is realised in the space $F$
of Theorem 1. According to this theorem
we have $H_{\xi_{F}\eta_{F}}=\oplus_{m=0}^{\infty}
d_{\xi_{F}\eta_{F}}^{m}
Ker a_{\xi_{F}\eta_{F}}$ and to construct  the representation
we have to study an action of $c_{\xi_{F}\eta_{F}}$ and
$c^{*}_{\xi_{F}\eta_{F}}$
on $Ker a_{\xi_{F}\eta_{F}}$. The operator $c_{\xi_{F}\eta_{F}}$
maps this closed subspace into itself and obey the condition of unitarity.
To prove Proposition 3 it sufficent to construct in the closed subspace
$Kera_{\xi_{F}\eta_{F}}$  an orthonormal basis
$\{h_{nMN}, n$ is integer$\}$
obeing the condition
\begin{equation}
c h_{nMN}= h_{n+1,MN}
\end{equation}
\begin{equation}
c^{*} h_{nMN}= h_{n-1,MN}
\end{equation}
Taking $e_{mnMN}=\frac{d^{m}h_{nMN}}{||d^{m}h_{nMN}||}$ we will
get an orthonormal basis
obeing conditions of Proposition~3.
Having in Hilbert spaces $F_{\xi_{F}}$ and $F_{\eta_{F}}$ orthonormal
basises $\{e_{mM},m=0,1,2,...\}$  and $\{e_{nN},n=0,1,2,...\}$ in Hilbert
space  $H_{\xi_{F}\eta_{F}}$ we have an orthonormal basis
$\{e_{mM}\otimes e_{nN},m,n=0,1,2,...\}$.
By acting by $$a_{\xi_{F}\eta_{F}}=
a_{\xi_{F}}\otimes a_{\eta_{F}}+b_{\xi_{F}}\otimes c_{\eta_{F}}$$
on arbitrary vector $x$
 for   $x=\sum_{m,n=0,MN}^{\infty}x^{mnMN}e_{mM}\otimes e_{nN} \in
Ker a_{\xi_{F}\eta_{F}}$ we have the following.
\begin{equation}\label{kera}
x^{m+1,n+1,MN}=\frac{q^{m+n+1}}
{\sqrt{(1-q^{2(m+1)})(1-q^{2(n+1)})}}\sum_{K,L}c_{\xi K}^{*M}c_{\eta L}^{N}
 x^{mnKL}
\end{equation}
Let us define unitary operators $C_{\xi}$ and $C_{\eta}$ in Hilbert spaces
$F_{\xi}$ and $F_{\eta}$ according to following formulas.
\begin{equation}\label{defccapxi}
C_{\xi}e_{mM}=\sum_{K}c_{\xi M}^{K}e_{mK}
\end{equation}
\begin{equation}\label{defccapeta}
C_{\eta}e_{nN}=\sum_{L}c_{\eta N}^{L}e_{nL}
\end{equation}

Solving the recurrent relation (\ref{kera}) one gets
 in $Ker a_{\xi_{F}\eta_{F}}$
an orthogonal basis  $\{g_{nMN},n$~is~integer$\}$
defined according to the following formulas:
\begin{equation}
g_{nMN}=\sum_{k=0}^{\infty}C^{*k}_{\xi}C^{k}_{\eta}e_{n+k,M}\otimes e_{kN}
(\frac{q^{n}}{1-q^{2}})^{k}q^{k^{2}}
\sqrt{\frac{[n]_{q^{2}}!}{[n+k]_{q^{2}}![k]_{q^{2}}!}}
\end{equation}
for $n \ge 0$ and
\begin{equation}
g_{nMN}=\sum_{k=0}^{\infty}C^{*k}_{\xi}C^{k}_{\eta}e_{kM}\otimes e_{k-n,N}
(\frac{q^{-n}}{1-q^{2}})^{k}q^{k^{2}}
\sqrt{\frac{[-n]_{q^{2}}!}{[-n+k]_{q^{2}}![k]_{q^{2}}!}}
\end{equation}
for $n<0$.

An operator $c_{\xi_{F}\eta_{F}}$ acts in this basis in the following way:
\begin{equation} \label{actc1}
c g_{n}=\frac{1}{\sqrt{1-q^{2(n+1)}}}C_{\eta}g_{n+1,MN}
\end{equation}
for $n \ge 0$ and

\begin{equation}\label{actc2}
c g_{nMN}= \sqrt{1-q^{-2n}}C_{\xi}g_{n+1MN}
\end{equation}
for $n<0$.

Let us prove for example the first formula (\ref{actc1}).
$$
c_{\xi\eta}C^{*k}_{\xi}C^{k}_{\eta}e_{n+k,M}\otimes e_{kN}
(\frac{q^{n}}{1-q^{2}})^{k}q^{k^{2}}
\sqrt{\frac{[n]_{q^{2}}!}{[n+k]_{q^{2}}![k]_{q^{2}}!}}
= $$
$$=\frac{1}{\sqrt{1-q^{2(n+1)}}}C_{\eta}(
C^{*(k-1)}_{\xi}C^{k-1}_{\eta}e_{n+k,M}\otimes e_{k-1,N}
(\frac{q^{n+1}}{1-q^{2}})^{k}q^{k^{2}}
\sqrt{\frac{[n+1]_{q^{2}}!}{[n+k]_{q^{2}}![k-1]_{q^{2}}!}}q^{2(n+k)}+
$$\begin{equation}
+C^{*k}_{\xi}C^{k}_{\eta}e_{n+1+k,M}\otimes e_{kN}
(\frac{q^{n+1}}{1-q^{2}})^{k}q^{k^{2}}
\sqrt{\frac{[n+1]_{q^{2}}!}{[n+1+k]_{q^{2}}![k]_{q^{2}}!}}(1-q^{2(n+k+1)})
)
\end{equation}
Summating over $k$ we will get the formula (\ref{actc1}).
In analoguous way one can get the formula (\ref{actc2}).

Taking
\begin{equation}
h_{nMN}=C_{\eta}^{n}\frac{g_{0MN}}{||g_{0MN}||}=\frac{1}
{([n]_{q^{2}}!(1-q^{2})^{n})^{\frac{1}{2}}}C_{\eta}^{n}
\frac{g_{nMN}}{||g_{0MN}||}
\end{equation}
for $n \ge 0$ and
\begin{equation}
h_{nMN}=C^{*(-n)}_{\xi}\frac{g_{0MN}}{||g_{0MN}||}=\frac{1}
{([-n]_{q^{2}}!(1-q^{2})^{-n})^{\frac{1}{2}}}C^{*(-n)}_{\xi}
\frac{g_{nMN}}{||g_{0MN}||}
\end{equation}
for $n<0$ we get demanding basis. This finishes the proof of Proposition 3.

Proving the Proposition 3 we get the following additional result:
using unitarity of $C_{\eta}$ and formulas (\ref{actc1}) and (\ref{actc2})
we get for $n\ge0$
$$||g_{n+1,MN}||^{2}=(1-q^{2})[n+1]_{q^{2}}||g_{nMN}||^{2}.$$
Let us  define
\begin{equation}
g_{q}(n)=\frac{||g_{nMN}||^{2}}{[n]_{q^{2}}!}=\sum_{k=0}^{\infty}
(\frac{q^{n}}{1-q^{2}})^{2k}q^{2k^{2}}
\frac{1}{[n+k]_{q^{2}}![k]_{q^{2}}!}
\end{equation}

{\bf Corollary 2. }{\it The defined series $g_{q}(n)$
 has the property of a character:  $$g_{q}(n+1)=(1-q^{2})g_{q}(n)$$.}

{\bf Proposition 4. }{\it
Representation $\xi_{F}\eta_{G}$ is realised in the Hilbert space
$F_{\xi}\otimes G_{\eta}$. This Hilbert space possess the orthonormal basis
$\{e_{mM\beta},m=0,1,2...\}$. In this basis
representation  $\xi_{F}\eta_{G}$
is defined according to the following formulas:
\begin{equation}
a e_{mM\beta}=\sqrt{1-q^{2m}}e_{m-1,M\beta} ; a e_{0M\beta}=0
\end{equation}
\begin{equation}
d e_{mM\beta}=\sqrt{1-q^{2(m+1)}}e_{m+1,M\beta}
\end{equation}
\begin{equation}
c e_{mM\beta}= q^{m}\sum_{K}c_{\xi M}^{K} e_{mK\beta}
\end{equation}
\begin{equation}
b e_{mM\beta}=- q^{m+1}\sum_{K}c_{\xi M}^{*K} e_{mK\beta}
\end{equation}
}

{\bf Proof}

According to Lemma 3 we have $Kerc_{\xi_{F}\eta_{G}}=0$.
Therefore this representation is realised in the space $F$
of Theorem 1. According to this theorem
we have $H_{\xi_{F}\eta_{G}}=\oplus_{m=0}^{\infty}
d_{\xi_{F}\eta_{G}}^{m}
Ker a_{\xi_{F}\eta_{G}}$ and to construct  the representation
we have to study an action of $c_{\xi_{F}\eta_{G}}$ and
$c^{*}_{\xi_{F}\eta_{G}}$
on $Ker a_{\xi_{F}\eta_{G}}$. The operator $c_{\xi_{F}\eta_{G}}$
maps this closed subspace into itself and obey the condition of unitarity.
To prove Proposition 4 it sufficent to construct in the closed subspace
$Kera_{\xi_{F}\eta_{G}}$  an orthonormal basis
$\{h_{M\beta}\}$
obeing the condition
\begin{equation}
c h_{M\beta}=\sum_{K}c_{\xi M}^{K} h_{K\beta}
\end{equation}
\begin{equation}
c^{*} h_{M\beta}=\sum_{K}c_{\xi M}^{*K} h_{K\beta}
\end{equation}
Taking $e_{mM\beta}=\frac{d^{m}h_{M\beta}}{||d^{m}h_{M\beta}||}$ we will
get an orthonormal basis
obeing conditions of Proposition~4.
Having in Hilbert spaces $F_{\xi_{F}}$ and $F_{\eta_{G}}$ orthonormal
basises $\{e_{mM},m=0,1,2,...\}$  and $\{e_{\beta}\}$ in Hilbert
space  $H_{\xi_{F}\eta_{G}}$ we have an orthonormal basis
$\{e_{mM}\otimes e_{\beta},m=0,1,2,...\}$.
By acting by $$a_{\xi_{F}\eta_{G}}=
a_{\xi_{F}}\otimes a_{\eta_{G}}+b_{\xi_{F}}\otimes c_{\eta_{G}}$$
on arbitrary vector $x$
 for   $x=\sum_{m=0,M\beta}^{\infty}x^{mM\beta}e_{mM}\otimes e_{\beta} \in
Ker a_{\xi_{F}\eta_{G}}$ we have the following:
\begin{equation}\label{kera4}
x^{mM\beta}=0
\end{equation}
for $m>0$.

Using the relation (\ref{kera4}) one gets
 in $Ker a_{\xi_{F}\eta_{G}}$
an orthonormal basis  $\{h_{M\beta}\}$
defined according to the following formulas:
\begin{equation}
h_{M\beta}=a_{\eta}^{M}e_{0M}\otimes e_{\beta}
\end{equation}
This basis satisfies conditions of representation 4.

In analogous way one can prove the proposition 5.

{\bf Proposition 5. }{\it
Representation $\xi_{G}\eta_{F}$ is realised in the Hilbert space
$G_{\xi}\otimes F_{\eta}$. This Hilbert space possess the orthonormal basis
$\{e_{\alpha nN},n=0,1,2...\}$. In this basis
representation  $\xi_{G}\eta_{F}$
is defined according to the following formulas:
\begin{equation}
a e_{\alpha nN}=\sqrt{1-q^{2n}}e_{\alpha,n-1,M} ; a e_{\alpha 0N}=0
\end{equation}
\begin{equation}
d e_{\alpha nN}=\sqrt{1-q^{2(n+1)}}e_{\alpha,n+1,N}
\end{equation}
\begin{equation}
c e_{\alpha nN}= q^{n}\sum_{L}c_{\eta N}^{L} e_{\alpha nL}
\end{equation}
\begin{equation}
b e_{\alpha nN}=- q^{n+1}\sum_{L}c_{\eta N}^{*L} e_{\alpha nL}
\end{equation}
}

The proof of the following proposition one can get using Theorem 1 and
Lemma 3.

{\bf Proposition 6. }{\it
Representation $\xi_{G}\eta_{G}$ is realised in the Hilbert space
$G_{\xi}\otimes G_{\eta}$. This Hilbert space possess the orthonormal basis
$\{e_{\alpha\beta},n=0,1,2...\}$. In this basis
representation  $\xi_{G}\eta_{G}$
is defined according to the following formulas:
\begin{equation}
a e_{\alpha\beta}=\sum_{\gamma\delta}a_{\xi\alpha}^{\gamma}
a_{\eta\beta}^{\delta}e_{\gamma\delta}
\end{equation}
\begin{equation}
d e_{\alpha\beta}=\sum_{\gamma\delta}a_{\xi\alpha}^{*\gamma}
a_{\eta\beta}^{*\delta}e_{\gamma\delta}
\end{equation}
\begin{equation}
c e_{\alpha\beta}=0
\end{equation}
\begin{equation}
b e_{\alpha\beta}=0
\end{equation}
}

Using Theorem 1 and propositions 3,4,5,6 one can construct coproduct
of arbitrary representations of $A$.

\section{ A construction of representation $\Pi$.}
In the present section we will use material of previous sections to construct
a stable representation $\Pi$ and to investigate its properties.
Using Propsition 3 we can get propositions 7 and 8.

{\bf Proposition 7. }{\it
Representation  $\pi_{\phi}\pi_{\psi}$ where $\pi_{\phi}$
is irreducible representation of infinitely dimensional family
is realised in the
Hilbert space $H_{\pi_{\phi}\pi_{\psi}}$ with orthonormal basis
$\{ e_{mM}\} $ where $ m $  runs from $ 0 $ to $ \infty$
and $ M $  runs on all integer numbers
according to the following formulas:
\begin{equation}
a e_{mM}=\sqrt{1-q^{2m}}e_{m-1,M} ; a e_{0M}=0
\end{equation}
\begin{equation}
d e_{mM}=\sqrt{1-q^{2(m+1)}}e_{m+1,M}
\end{equation}
\begin{equation}
c e_{mM}= q^{m} e_{m,M+1}
\end{equation}
\begin{equation}
b e_{mM}=- q^{m+1} e_{m,M-1}
\end{equation}
}

Let us construct the representation $\Pi=\pi_{\phi}\pi_{\psi}\pi_{\theta}$.

{\bf Proposition 8. }{\it
Representation  $\Pi=\pi_{\phi}\pi_{\psi}\pi_{\theta}$ is realised in the
Hilbert space $H_{\Pi}$ with orthonormal basis
$\{ e_{mM\Gamma}\} $ where $m$  runs from $0$ to $\infty$
and $ M$ and $\Gamma $  runs on all integer numbers
according to the following formulas:
\begin{equation}
a e_{mM\Gamma}=\sqrt{1-q^{2m}}e_{m-1,M\Gamma} ; a e_{0M\Gamma}=0
\end{equation}
\begin{equation}
d e_{mM\Gamma}=\sqrt{1-q^{2(m+1)}}e_{m+1,M\Gamma}
\end{equation}
\begin{equation}
c e_{mM\Gamma}= q^{m} e_{m,M+1,\Gamma}
\end{equation}
\begin{equation}
b e_{mM\Gamma}=- q^{m+1} e_{m,M-1,\Gamma}
\end{equation}
}

Let us investigate properties of representation $\Pi$. Applying the
Proposition 2 we get the following proposition.

{\bf Proposition 9. }{\it
A weak closure of $A$ in representation $\Pi$
is a factor of a type II${}_{\infty}$.
A trace on this factor is connected with   the
invariant integral of Woronowich on the quantum group
 $SU_{q}(2)$  by the formula $\int f d\mu  = tr(f cc^{*})$.

}

 {\bf Proof}

According to Proposition 2 the representation $\Pi$ is factorial of a type
II$_{\infty}$. According to Proposition 2 the trace on $A$
in the representation  $\Pi$ will takes the following form.
\begin{equation}
tr(f)=\sum_{k=0}^{\infty} f_{k00}^{k00}
\end{equation}

This trace is connected with the invariant integral
of Woronovich on $SU_{q}(2)$  through the formula  $\int f d\mu  = tr(f
cc^{*})$ up to multiplicative constant.

{\bf Proposition 10.}{\it The direct sum of equivalent to $\Pi$ representations
is equivalent to $\Pi$.

}

In more formal view the Proposition 10 looks as follows.
Let $\Pi$ be realised in the Hilbert space $H_{\Pi}$ and let $H$
be arbitrary Hilbert space.

Let us consider in the Hilbert space $H_{\Pi}\otimes H$ representation
$\Xi$ defined according to the following formula.
$$\Xi(f)=\Pi(f)\otimes Id_{H}$$
for $f\in A$. According to  Proposition 10  the representation $\Xi$
is unitary equivalent to $\Pi$.

Let us investigate the property of stability for
representation $\Pi$.

{\bf Theorem 2. }{\it
For arbitrary *-representation  $\xi$ of $A$
the following condition of stability does hold:
$$\Pi\xi=\xi\Pi=\Pi$$
where "=" means unitary equivalence.
}

{\bf Proof}

Let us first prove that $\Pi\xi=\Pi$.

According to Theorem 1 and Lemma 2  we have the decomposition of
the representation $\xi$ in the direct sum.
$$\xi=\xi_{F}\oplus\xi_{G}$$
Here representations
$\xi_{F}$ and $\xi_{G}$ are representations defined in the previos section.
It sufficient to prove that
$$\Pi\xi_{F}=\Pi$$
$$\Pi\xi_{G}=\Pi.$$
To prove these conditions we use propositions 3,4 and 10. Applying
the Proposition 10 we get condition $\Pi\xi=\Pi$.

Analogously one can prove condition $\xi\Pi=\Pi$.
This finishes the proof of the Theorem 2.

\section{ The second property of stability for representation $\Pi$.}
In this section we will investigate the second property of stability
for representation $\Pi$. Let us consider representation $\Pi$ in the
following form.
Representation  $\Pi$ is realised in the
Hilbert space $H_{\Pi}$ with orthonormal basis
$\{ e_{mM\Gamma}\} $ where $m$ and $\Gamma$ runs from $0$ to $\infty$
and $M$  runs on all integer numbers
according to the following formulas:
\begin{equation}
a e_{mM\Gamma}=\sqrt{1-q^{2m}}e_{m-1,M\Gamma} ; a e_{0M\Gamma}=0
\end{equation}
\begin{equation}
d e_{mM\Gamma}=\sqrt{1-q^{2(m+1)}}e_{m+1,M\Gamma}
\end{equation}
\begin{equation}
c e_{mM\Gamma}= q^{m} e_{m,M+1,\Gamma}
\end{equation}
\begin{equation}
b e_{mM\Gamma}=- q^{m+1} e_{m,M-1,\Gamma}
\end{equation}
Let us define in the same Hilbert space $H_{\Pi}$ representation of $A$
$\Pi'$ generated by operators $a',b',c',d'$ according the following formulas.
\begin{equation}
a' e_{mM\Gamma}=\sqrt{1-q^{2\Gamma}}e_{mM,\Gamma-1} ; a e_{mM0}=0
\end{equation}
\begin{equation}
d' e_{mM\Gamma}=\sqrt{1-q^{2(\Gamma+1)}}e_{mM,\Gamma+1}
\end{equation}
\begin{equation}
c' e_{mM\Gamma}= q^{\Gamma} e_{m,M+1,\Gamma}
\end{equation}
\begin{equation}
b' e_{mM\Gamma}=- q^{\Gamma+1} e_{m,M-1,\Gamma}
\end{equation}
One can prove that the representation $\Pi'$ is unitary equivalent to $\Pi$.
It is easy to see that operators $a',b',c',d'$  commute with
operators $a,b,c,d$. In fact we can prove the following proposition.

{\bf Proposition 11.}{\it The weak closure of $\Pi'(A)$ is equal to
commutant of $\Pi(A)$.}

According to Proposition 2
we get two factors of a type II$_{\infty}$ that are commutants
of each other. These factors are weak closures of $\Pi(A)$ and $\Pi'(A)$.

Let us define the representation $\Pi''(A)$ according to the following formula.

\begin{equation}
\left(\begin{array}{cc}a''&b''\\c''&d''\end{array}\right)=
\left(\begin{array}{cc}a&b\\c&d\end{array}\right)
\left(\begin{array}{cc}a'&b'\\c'&d'\end{array}\right)
\end{equation}

Using calculations close to considered in the previos sections one can
prove the following theorem.

{\bf Theorem 3.}{\it Representations $\Pi"$ is unitary equivalent to $\Pi$.}

Let us call this property the second property of stability for
representation $\Pi$.

\section{Conclusion.}
Some analog of a stable representation of a bialgebra was considered
in the work [6] of Faddeev and Takhtadjan. They used some kind of
stable representation
for calculation of a monodromy matrix in some statistical mechanic problems.

Representation $\Pi$ can be useful for construction of evolution on the
quantum group. In the work [7] of Arefeva, Parthasarathy, Visvanathan and
Volovich
it were considered some examples of evolution on the quantum group.
It is natural to obtain the evolution of a point on a group as a product of
matrixes.
For quantum groups this can be formulated as a coproduct on a quantum group
function algebra. If we want to get the evolution of a quantum group function
algebra  expressed by an evolution operator we should have the representation
of a quantum group function algebra invariant under coproduct.
Representation $\Pi$ satisfy this condition for coproduct by any irredusible
representation.

In fact representations considered in this paper were used in
the work  [8] of Arefeva and Arutyunov  about representation of a q-deformed
de Rama complex on a quantum group. They used representation of
$SU_{q}(2)$ function algebra that is equivalent to $\pi_{\phi}\pi_{\psi}$.
In their work this representation can be replaced by $\Pi$.

Author is grateful to I.V.Volovich and E.I.Zelenov for discussions.


\begin{thebibliography}{99}
\bibitem{Drn} \ Drinfeld V.D. ZNS LOMI 155(1986)p.19
\bibitem{Wor} \ Woronovich S.L.  Comm.Math.Phys. 111(1987)N4,p.613,
Publ.RIMS Kyoto Univ. 23(1986)p.117
\bibitem{FRT} \ Faddeev L.D., Reshetikhin N.Yu., Takhtadjan L.A.
Alg.Anal.1(1988)p.129
\bibitem{Jap} \ Masuda T., Mimachi K.,  Nakagami Y., Noumi M., Ueno K.
J.Func.Anal 99(1991)N2
\bibitem{VaS} \ Vaxman L.L., Soibelman Ya.E. Func.Anal.Appl.22(1988)N3,p.1
\bibitem{FaT} \ Faddeev L.D.,  Takhtadjan L.A. Usp.Math.Nauk.34(1979)N5
\bibitem{APVV} \ Coherent States, Dynamics and Semiclassical Limit
on Quantum Groups.
Arefeva I.Ya., Parthasarathy R., Viswanathan K.S., Volovich I.V.
preprint SFU-HEP-108-93
\bibitem{ArA} \ Uniqueness of $U_{q}(N)$ as a quantum gauge group and
representations of its differential algebra.
 Arefeva I.Ya., Arutyunov G.E. preprint SMI-8-93
\bibitem{Sak} \ Sakai S. $C^{*}$-algebras and $W^{*}$-algebras
Springer-Verlag 1971
\end{thebibliography}
\end{document}